\documentclass[prl,twocolumn,amssymb]{revtex4}
\usepackage{epsfig}
\begin{document}
\title[Induced Magnetic Ordering]{Induced
Magnetic Ordering by Proton Irradiation in Graphite}
\author{P. Esquinazi}\email[E-mail address: ]{esquin@physik.uni-leipzig.de}
\author{D. Spemann}\author{R. H\"ohne}\author{A. Setzer}\author{K.-H. Han}
\author{T. Butz} \affiliation{Institut f\"ur Experimentelle Physik
II, Universit\"at Leipzig, Linn{\'e}str. 5, D-04103 Leipzig, Germany}

\begin{abstract}
We provide evidence that proton irradiation of energy 2.25 MeV  on
highly-oriented pyrolytic graphite samples triggers ferro- or
ferrimagnetism. Measurements performed with a superconducting quantum
interferometer device (SQUID) and magnetic force microscopy (MFM) reveal
that the magnetic ordering is stable at room temperature.
\end{abstract}

\maketitle

The search for macroscopic magnetic ordering phenomena in organic
materials has been pushed to the frontiers of physics, chemistry and
materials science since the discovery of ferromagnetism in the open-shell
radical $p$-nitrophenylnitronyl nitroxide ($p-$NPNN) \cite{turek} and in
tetrakis(dimethylamino)ethylene [TDAE]$^+$C$_{60}$ charge-transfer salt
\cite{alle} with Curie temperatures $T_c = 0.6~$K and 16~K, respectively.
Although a remarkable amount of studies have been performed in the last
years building molecules with unpaired $\pi$-electrons \cite{pi}, the
scientific community did not become aware of the possible existence of a
room temperature metal-free organic ferromagnet till the accidental
``discovery" of ferromagnetism in polymerized fullerenes was reported
\cite{maka,wood}. The study of the magnetic properties of polymerized
fullerene was triggered by the previously reported weak ferromagnetic-like
magnetization loops in highly-oriented pyrolytic graphite (HOPG)
\cite{yakov} that indicate the existence of magnetic ordering with $T_c$
much above room temperature \cite{esqui}. The origin for the reported
ferro- or ferrimagnetism in polymerized C$_{60}$ as well as in HOPG
remains unclear. The contribution of ferromagnetic impurities, like iron,
has been thoroughly investigated \cite{esqui,hohne} and it is something
one should carefully take into account, specially when the ferromagnetic
signal under study is small. Ferromagnetism in a graphite nodule from a
meteorite with $T_c \simeq 570~$K has been recently suggested to be due to
a large magnetic proximity effect at the interface with magnetite or
kamacite inclusions \cite{coey}. As in Ref.~\cite{coey} most of the
magnetism specialists  doubt that a room-temperature magnet can exist in a
material with only s- and p-electrons. Clear evidence for its existence
and reproducibility as well as the report of a relatively easy method to
produce it, is of fundamental interest for a broad spectrum of natural
sciences.

Early reports on room-temperature ferromagnetic behavior in some
carbon-based structures were apparently not taken seriously by the
scientific community, partially due to the weakness of the ferromagnetic
signals added to the unknown contribution of magnetic impurities and low
reproducibility \cite{makarev}. From those early works our attention was
focused to the  magnetic properties found in amorphous-like carbon
prepared from different hydrogen-rich starting materials where an increase
of the saturation magnetization with the hydrogen concentration in the
starting material was found \cite{murata}. The origin for the magnetic
ordering has been related to the mixture of carbon atoms with sp$^2$ and
sp$^3$ bonds, which was predicted to reach a magnetization higher than in
pure Fe \cite{ovchi}. Hydrogen, on the other hand, was assumed to have a
role only in the formation of the amorphous carbon structure
\cite{murata}. New theoretical predictions, however, show that
hydrogenated graphite can display spontaneous magnetization coming from
different numbers of mono- and dihydrogenated carbon atoms \cite{kusa}. We
show in this letter that implanted protons in HOPG triggers ferromagnetic
(or ferrimagnetic) ordering with a Curie temperature above room
temperature.

Two HOPG (ZYA grade) samples from Advanced Ceramics Co. with a content of
magnetic metallic impurities below 1 ppm \cite{esqui} were irradiated with
a 2.25 MeV proton microbeam. The samples' dimension was $2 \times 2 \times
0.1~$mm$^3$. Before irradiation they were glued with varnish on a
high-purity Si substrate and the magnetic moment of the whole ensemble as
well as of the Si substrate alone were measured. Due to the small sample
mass the contribution of the HOPG virgin sample ferromagnetic signal
(saturation magnetization $M_s \sim 3 \times 10^{-4}~$emu/g at
temperatures $T \le 300~$K \cite{esqui}) is overwhelmed by
 the diamagnetic signal of the
Si-substrate for fields applied parallel to the graphene layers.
Rutherford Backscattering Spectroscopy and Particle Induced X-ray Emission
spectra were recorded simultaneously with the irradiation allowing us to
check the purity of the sample at the different irradiation stages.
Assuming that all the (measured) Fe impurities  ($< 0.3 \mu$g Fe per gram
graphite) in our samples behave as bulk ferromagnetic material, a rather
unrealistic assumption, their maximum contribution to the magnetic moment
in our samples would be $m < 6.1 \times 10^{-8}$~emu. The irradiated and
non-irradiated areas were further characterized by atomic force and
magnetic force microscopy (AFM,MFM) with a Nanoscope III scanning probe
microscope operating in ``tapping/lift" mode, using standard tips coated
with a magnetic CoCr film magnetized normal to the sample surface. The
magnetization measurements were performed with a SQUID magnetometer from
Quantum Design with the reciprocating sample option (RSO) and a
sensitivity of $\lesssim 10^{-7}$emu.

Four irradiations were consecutively applied to sample 1, namely: stage
\#1: homogeneous irradiation of an area $1720 \times 1720~\mu$m$^2$, dose:
0.99 pC/$\mu$m$^2$, total charge: 2.93 $\mu$C; \#2: $100 \times 100$ spots
of 2$\mu$m diameter each, on an area $570 \times 570~\mu$m$^2$ in the
middle of the sample, dose: 0.3 nC/$\mu$m$^2$, total charge: $\simeq
8~\mu$C; \#3: 4 spots of 0.8~mm diameter each, dose: 0.3~nC/$\mu$m$^2$,
total charge: $\simeq 600~\mu$C; \#4: the same as \#3. To check the
reproducibility of our procedure as well as to rule out possible
contamination during the handling of the sample, a new piece of the virgin
HOPG sample was prepared in a similar way and fixed to a different
Si-substrate. On this sample~2 three spots of 0.8~mm diameter, dose:
0.3~nC/$\mu$m$^2$, total charge: 450~$\mu$C were irradiated, similar to
stage \#3 but one spot less.

To estimate the defect density created by the proton beam, Monte Carlo
simulations (SRIM2003 \cite{zie}) were performed with full damage cascades
assuming a displacement energy of $E_d = 35~$eV for the creation of a
Frenkel pair in graphite. According to these calculations, 2.25 MeV
protons come to rest at a depth of $\sim 46~\mu$m from the graphite
surface with a
 $1.75~\mu$m width (full width at half maximum of the
implantation profile). Since hydrogen is fully trapped in HOPG up to a
concentration of 0.45 H/C \cite{sie}, all implanted protons will be
trapped  for doses up to 14 nC/$\mu$m$^2$ when they come to rest in the
HOPG sample. Assuming that no annealing of defects during irradiation
occurs and that the damaged regions are of the size of the irradiated one,
we obtain that for proton dose of 0.3~nC/$\mu$m$^2$ the defect density is
$\sim 2 \times 10^{20}~$cm$^3$, i.e. only $\sim 0.1\%$ of the carbon atoms
are displaced. For a dose 250 times higher we expect that 35\% of the
carbon atoms are displaced and the carbon lattice is amorphous.

\begin{figure}
\centerline{\psfig{file=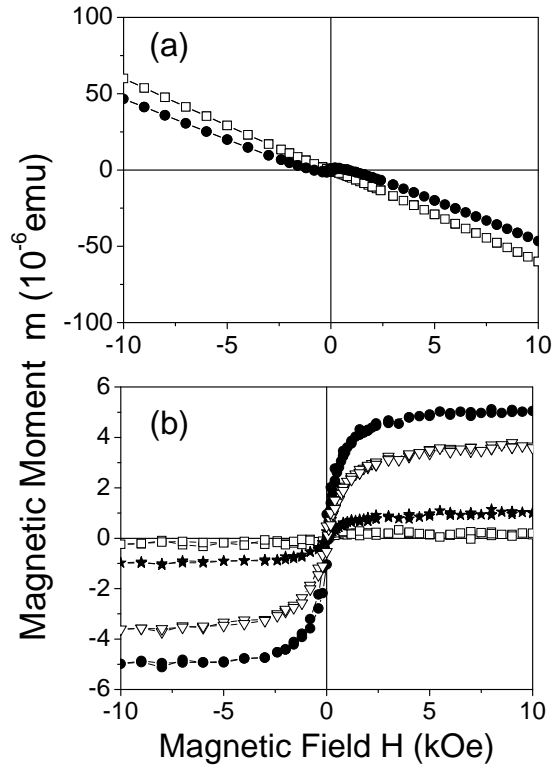,width=8.0cm}} \caption{The magnetic
moment (in units of $10^{-6}$emu $= 10^{-9}$Am$^2$) measured at $T = 300$K
as a function of magnetic field (1kOe $= 10^6/4\pi$Am$^{-1}$) by cycling
the field from zero to +10~kOe, from +10~kOe to -10~kOe and back to
+10~kOe for sample~1 glued on a silicon substrate, after various proton
irradiations. (a) Total magnetic moment without any background subtraction
for the sample after stage \#1 ($\Box$) and \#3 ($\bullet$) irradiations.
(b) Magnetic moment after subtraction of the sample holder magnetic
moment, for the sample after the first ($\Box$), second ($\star$), third
($\bullet$) and fourth ($\bigtriangledown$) irradiation stages.}
\label{sample1}
\end{figure}

Figure~\ref{sample1}(a) shows the measured magnetic moment of sample 1
after irradiation \#1, which is similar to the virgin sample within the
scale of the figure. The field was applied parallel to the graphene
planes. The main part ($\sim 90\%$) of the diamagnetic signal is due to
the Si substrate. In the same figure we show the magnetic moment of the
same ensemble after irradiation \#3 where we can clearly recognize the
s-shape curve, see Fig.~\ref{sample1}(a), without any background
substraction. After subtraction of the magnetic moment of the substrate we
obtain for sample~1 the results depicted in Fig.~\ref{sample1}(b). We
observe a clear increase of the ferromagnetic loop with irradiation steps
\#2 and \#3. For the last step \#4 the magnetic moment of the irradiated
sample decreased slightly. The coercive fields are between 80 and 110~Oe
and weak temperature dependent. The remanent magnetic moments are
approximately 20\% of the values of the saturation magnetic moments. For
the sample with the highest magnetic moment (sample 1 after stage \#3):
$m_{\rm rem} \simeq 1 \times 10^{-6}$emu.

The observed changes of the magnetic moment after the different
irradiation stages are due to the irradiation and not to different
misalignments of the sample position respect to the applied field. To
check this we measured sample~1 in the irradiation stage \#3 and \#4 for
the other field direction (parallel to the c-axis of graphite). After
subtraction of the diamagnetic signal from graphite and Si substrate, the
magnetic moment loops were similar as for the other field direction; this
is an indication for a low anisotropy of the magnetic ordering produced by
the irradiation.

\begin{figure}
\centerline{\psfig{file=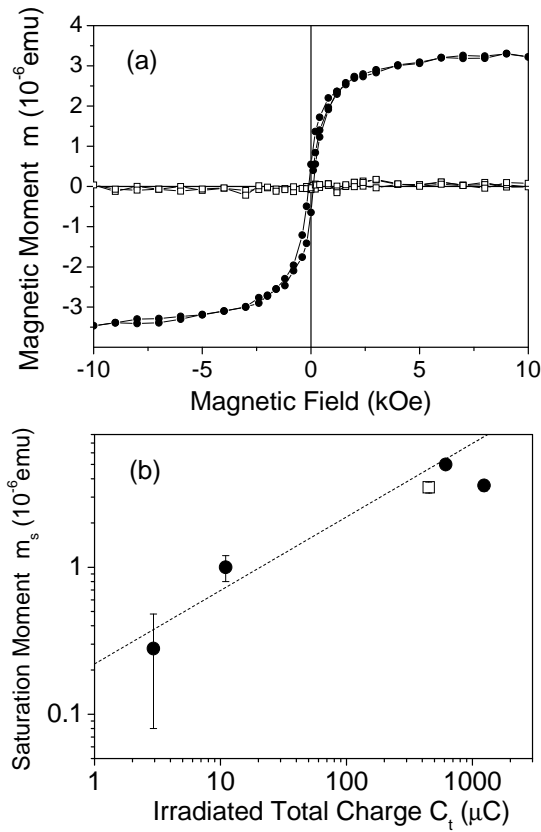,width=8.0cm}} \caption{(a) Magnetic
moment (in units of $10^{-6}$emu) measured at $T = 300$K as a function of
magnetic field  as in Fig.1(b), for sample~2 before ($\Box$) and after
($\bullet$) proton irradiation. (b) Measured saturation magnetic moment as
a function of the total irradiated charge $C_t$ for sample~1 ($\bullet$)
and for sample~2 ($\Box$). The dashed line is the funtion $m_s =
0.22[10^{-6}$emu$/\mu$C$^{0.5}] C_t^{0.5}$.} \label{sample2}
\end{figure}

Figure~\ref{sample2}(a) shows the magnetic moment of sample 2 before and
after irradiation of the three spots. The background signal due to the
substrate was subtracted. We see a clear development of the ferromagnetic
signal after irradiation. The saturation magnetic moment is slightly
smaller than the one we would expect assuming a linear relation between
saturation moment $m_s$ and total implanted charge $C_t$, namely  $\sim
(3/4) \times 5 \times 10^{-6} = 3.8 \times 10^{-6}~$emu. This value is
$10\%$ larger than the measured value $m_s \simeq 3.5 \times 10^{-6}~$emu.
Although the number of points is too small to provide with certainty the
relationship between $m_s$ and $C_t$, our results indicate a relation of
the type $m_s \propto \sqrt{C_t}$, see Fig.~\ref{sample2}(b). It is
interesting to note that after the \#4 irradiation the magnetic moment of
sample 1 decreases, see Fig.~\ref{sample1}(a), indicating that there might
be a competition between the produced disorder and the implanted charge,
which determines the total magnetic ordering.

Hysteresis loops were measured at 5 K, 300 K and 380 K. In this
temperature range there is no significant change of the ferromagnetic
loops with temperature. This result is similar to that obtained for the
``intrinsic" ferromagnetic-like loops of graphite in Ref.~\cite{esqui},
and indicates a Curie temperature much above 400~K, which is of importance
for future applications.

\begin{figure}
\centerline{\psfig{file=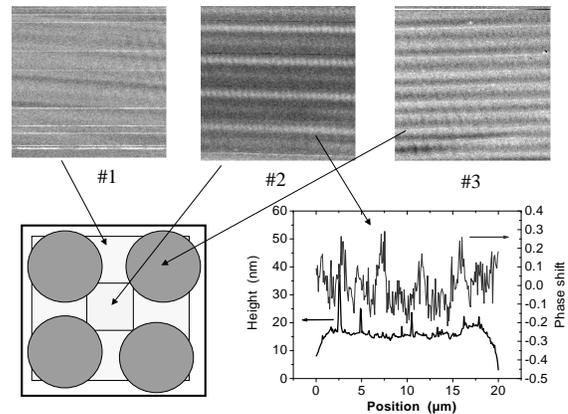,width=9.0cm}} \caption{Top: Phase
gradient pictures obtained from MFM at three surfaces of sample~1
corresponding to the irradiation stages \#1, \#2 and \#3, see sketch at
the bottom left of the figure. Bottom right: Topography and phase gradient
line scans of the middle MFM picture obtained at stage \#2.} \label{mfm}
\end{figure}

The MFM pictures obtained at the irradiated areas show
homogeneously distributed magnetic domains, as depicted in
Fig.~\ref{mfm}. The periodicity and/or the width of the domains
depends on the irradiation dose, ranging from 4 $\mu$m in the
second irradiated stage area to about $2 \mu$m at the area of the
third stage. A weak domain structure is recognized at the area of
 stage \#1, see Fig.~\ref{mfm}. We stress that the magnetic
signal is not correlated to the topography in any irradiation stage.  For
comparison, in Fig.~\ref{mfm} we show a topography and magnetic line-scans
obtained normal to the magnetic domain structure at stage \#2. The MFM
pictures shown in Fig.~\ref{mfm} were obtained at a distance of 50~nm
between tip and surface. Due to the small coercive field of the magnetic
surface and the influence of the magnetic tip we have observed that the
magnetic domain distribution depends on the distance between tip and
surface. Theoretically, the phase shift $\Delta \phi$ measured by MFM
should be proportional to the magnetic force gradient, which should depend
on the distance between the tip apex and sample surface $h$ as $(h +
\delta)^{-5}$, being $\delta$ the distance between tip apex and the
effective position of the magnetic moment in the tip. Measuring $\Delta
\phi$ as a function of $h$ we estimate $\delta \sim (900 \pm 100)$~nm.
With this value we estimate the maximum magnetic moment observed by the
tip $m \sim 3 \times 10^{-15}~$Am$^2$ at the area of largest magnetic
moment (stages \#3 and \#4). We note that measurements on different
magnetic samples indicate that $\delta$ depends on the magnetic properties
of the measured surface and therefore the calculated magnetic moment from
MFM data should be taken as a rough estimate only \cite{han}. Assuming
that the magnetic moment measured by MFM corresponds to a region of area
$\sim 10^{-12}$~m$^2$ and $\sim 1 \mu$m depth, we can estimate the total
saturation magnetic moment $m_s$ the SQUID would measure at stage $\#3$ as
$m_s \sim [3 \times 10^{-15} \times 4 \times \pi \times (0.4\times
10^{-3})^2 / 10^{-12}] 10^3$~emu$ = 6 \times 10^{-6}~$emu, in reasonable
agreement with the SQUID measurements.

Although a proton penetration depth of $\sim 46~\mu$m is expected, we do
not know how large is the depth for the ferromagnetic active part of the
irradiated sample. This will be studied by irradiating samples of
different thickness in the future. If we assume a depth of $1~\mu$m we get
a magnetization $M \simeq 1.1~$emu/g for sample~1 after stage \#3, which
is  $\sim 1\%$ of that of magnetite and of the same order as that obtained
for amorphous-like carbon prepared by CVD from hydrogen-rich targets
\cite{murata}. From MFM, the maximum phase shift observed in the
irradiated regions is of the order of $\Delta \phi_{\rm max} \sim
0.3^\circ$. From measurements on regions with Fe impurities (mainly
Fe$_3$O$_4$ of micrometer size) with the same tip and arrangement
\cite{han} we obtain $\Delta \phi_{\rm max} \sim 10^\circ \ldots
20^\circ$. That means that $\Delta \phi_{\rm max}$ obtained in the
irradiated regions \#3 is of the order of 1\% of that of magnetite, in
agreement with the rough estimate made above using the SQUID data and
assuming 1~$\mu$m ferromagnetic active thickness.

Magnetic force microscopy measurements on proton-irradiated spots on HOPG
show that it is possible to create ferromagnetic active areas of $\sim
1~\mu$m$^2$ with very sharp borders of the magnetic force gradient,
created mainly at low proton doses \cite{hanp}. That means that it is
possible to produce magnetic microstructures of arbitrary shape by a
dedicated proton beam scanning system. To check the importance of
hydrogen, we have produced spots of similar topography with 1.5 MeV helium
ion irradiation at different doses ($> 0.05$~nC/$\mu$m$^2$). Those
irradiated regions showed no significant magnetic signal in MFM and stress
the significance of hydrogen in the formation of the magnetic ordering in
graphite \cite{hanp}.

From SQUID and MFM measurements we conclude that magnetic ordering appears
after proton irradiation in graphite and that this is stable at room
temperature. Our results provide a possible answer to the origin of
ferromagnetism in other carbon-based structures \cite{maka,hohne,coey},
which is not correlated with magnetic impurities. Future experiments
should clarify the lattice position of hydrogen in graphite as well as its
role in the magnetic ordering. The characterization  of our samples with
Micro-Raman as well as XPS is of particular interest and will be performed
in the future. Systematic investigations of the magnetic properties of
different carbon materials irradiated at different ion doses are necessary
to gain more insight into the physical processes involved as well as to
broader the spectrum for future applications. MFM measurements on proton
irradiated amorphous-carbon films indicate the existence of weak magnetic
ordering. As possible explanation for our results we refer to the work in
Ref.~\cite{kusa} where a sp$^3$-sp$^2$ ferrimagnetic structure can arise
in a graphene plane through the existence of mono- and dihydrogenated
carbon atoms. Future work should provide an answer to the question to what
extent the graphite structure represents a quantitative advantage for the
formation of the magnetic ordering.

Our experimental findings open up a new research field in magnetism with
possible applications in spin electronics, since according to
Ref.~\cite{kusa} the electrons in the hydrogenated graphene should have a
very large spin polarization. If the interpretation of hydrogen-induced
ferrimagnetism in HOPG \cite{kusa} is correct, then with our technique it
should be possible to create the smallest controlled magnet ever possible
by implanting hydrogen at the edges of carbon nanotubes.

\begin{acknowledgments}
One of the authors (P.E.) thanks K. Kusakabe for correspondence.
 This work is supported by DFG Grant ES 86/6-3.
\end{acknowledgments}

\end{document}